# Comparison of linear and non-linear soft tissue models with post-operative CT scan in maxillofacial surgery

M. Chabanas[1,3], Y. Payan[1], Ch. Marécaux[1,2], P. Swider[3] and F. Boutault[2]

1. Laboratoire TIMC-IMAG, CNRS UMR 5525, Université J. Fourier – Grenoble
Institut d'Ingénierie de l'Information de Santé (In3S) - 38706 La Tronche cedex, France

2. Service de chirurgie maxillo-faciale et plastique de la face, Hôpital Purpan – Toulouse
Place Baylac BP 3103, 31059 Toulouse Cedex 3, France

3. Laboratoire de Biomécanique, EA 3697, Université P. Sabatier, Hôpital Purpan – Toulouse
Amphithéâtre Laporte, Place Baylac BP 3103 - 31059 Toulouse cedex 3, France

Matthieu.Chabanas@imag.fr

**Abstract**

A Finite Element model of the face soft tissue is proposed to simulate the morphological outcomes of maxillofacial surgery. Three modelling options are implemented: a linear elastic model with small and large deformation hypothesis, and an hyperelastic Mooney-Rivlin model. An evaluation procedure based on a qualitative and quantitative comparison of the simulations with a post-operative CT scan is detailed. It is then applied to one clinical case to evaluate the differences between the three models, and with the actual patient morphology. First results shows in particular that for a "simple" clinical procedure where stress is less than 20%, a linear model seams sufficient for a correct modelling.

## 1. Introduction

Modeling the human soft tissue is of growing interest in medical and computer science fields, with a wide range of applications such as physiological analysis, surgery planning, or interactive simulation for training purpose (Delingette 1998). In maxillofacial surgery, the correction of face dismorphosis is addressed by surgical repositioning of bone segments (e.g. the mandible, maxilla or zygomatic bone). A model of the patient face to simulate the morphological modifications following bone repositioning could greatly improve the planning of the intervention, for both the surgeon and the patient.

Different models were proposed in the literature. After testing discrete mass-springs structures (Teschner et al. 1999), most of the authors used the Finite Element method to resolve the mechanical equations describing the soft tissue behavior. Keeve et al. (1998), Koch et al (1999) and Zachow et al. (2000) first developed linear elastic model. With more complex models, Gladilin et al. (2003) discussed the advantages of non-linear hypotheses, and Vandewalle et al. (2003) began accounting for tissue growth in their simulation.

One of the most important issue in soft tissue modeling is to assess the quality of the simulations. From a modeling point of view, it enables to evaluate and compare different methods, for example linear versus a non-linear models. This is above all essential for the surgeon since the use of a soft tissue model in actual surgical practice cannot be considered without an extensive clinical validation. While many models were proposed in the literature, few works propose satisfying validation procedures.

In this paper, we first propose different modeling hypotheses of the face soft tissue. An evaluation procedure is then detailed, based on a qualitative and quantitative comparison of the simulations with a post-operative CT scan. Results are presented for a clinical case of retromandibular correction. The gesture actually realized during the intervention is measured with accuracy, then simulated using the biomechanical model. Simulations are thus compared to assess the influence of modeling options, and their relevancy with respect to the real post-operative aspect of the patient.

## 2. Modeling the face soft tissue

A project for computer-aided maxillofacial surgery has been developed for several years in the TIMC (Grenoble, France) laboratory, in collaboration with the Purpan Hospital of Toulouse, France. In that context, a Finite Element model of the face soft tissue has been developed to simulate the morphological modifications resulting from bones repositioning. In Chabanas et al 2003, we mainly presented our methodology to generate patient-specific Finite Element models. A generic mesh was built, organized in two layers of hexahedrons and wedges elements. The principle was then to conform this generic model to the morphology of each patient. Using elastic registration, nodes of the mesh were non-rigidly displaced to fit the skin and skull surfaces of the patient reconstructed from a pre-operative CT scan.

Once a mesh of the patient is available, biomechanical hypothesis must be chosen to model the mechanical behavior of the face tissue. Three different methods are compared in this paper: a linear elastic model, under small then large deformation hypothesis, and an hyperelastic model.

### 2.1 Linear elastic model

A first hypothesis is to model the tissue as a homogeneous, linear elastic material. This assumption, which considers the stress/strain relationship of the system as always linear during the simulation, is called *mechanical linearity*. Although biological tissues are much more complex, this behavior was found coherent for a relative strain under 10 to 15% (Fung, 1993). The material properties can be described using the Hooke's law and just two rheological parameters, the Young modulus and the Poisson ratio.

A second option of modeling depends on the deformations range. In *small deformations* hypothesis (also named *geometrical linearity*), the Green–Lagrange formula linking the stress and strain tensors is linearized by neglecting the second order term (Zienkiewicz and Taylor, 1989). As a consequence, the formulation can be written as a linear matrix inversion problem, which is straightforward and fast to resolve. Under the *large deformations* hypothesis, the second order term is not neglected, which leads to a more accurate approximation but dramatically increases the computation complexity.

A linear material with small deformations is the most widely used hypothesis in the literature of facial tissue deformation modeling. Despite the fact it is probably limited due to the complexity of the tissue properties and the simulated surgical procedures, this model is certainly the first to be tested and compared with actual data. We had therefore implemented such a model, with rheological parameters of 15kPa for the Young modulus, and a 0.49 Poisson ratio (quasi-incompressibility). Simulations were carried out under both small and large deformations hypotheses.

## 2.2 Hyperelastic model

As stated in different papers, linear models become inaccurate when the displacements or the deformations are large (Picinbono et al. 2000, Gladilin et al. 2003). Numerical errors appears due to the non-invariance in rotation. A major shortcoming especially lies on the material constitutive law. Experiments on biological tissue (Fung 1993) have shown that the stress increase much faster than the stress as soon as the small deformation context is not applicable. This increase of the stiffness must be taken into account, which is not possible with a linear constitutive law such as the Hooke's law. While non-linear relations can be used with an elastic model, another modeling framework was preferred, the hyperelasticity, to directly account for all the non linearities (mechanical and geometrical) in the mathematical formulation.Whereas a material is said to be *elastic* when the stress $S$ at a point $X$ depends only on the values of the deformation gradient $F$, the material is said to be *hyperelastic* when the stress can be derived from the deformation gradient and from a stored strain energy function $W$:

$$S = \frac{\partial W}{\partial E}, \text{ where } E \text{ is the Lagrangian strain tensor.}$$

The strain energy $W$ is a function of multidimensional interactions described by the nine components of $F$. It is very difficult to perform experiments to determine these interactions for any particular elastic material. Therefore, various assumptions have been made to derive simplified and realistic strain energy functions. One of this assumption is the Mooney-Rivlin materials modelling (Mooney 1940). For this, the energy function W can be approximated by a 5 coefficients Mooney-Rivlin material, so that:

$$W = a_{10}(I_1 - 3) + a_{20}(I_1 - 3)^2 + a_{01}(I_2 - 3) + a_{02}(I_2 - 3)^2 + a_{11}(I_1 - 3)(I_2 - 3)$$

where $I_1$ and $I_2$ are the first and the second invariant of the deformation tensor E. Assuming a constitutive law for facial tissues that is close to the constitutive law proposed by Gerard et al. (2003) for the human tongue, a two parameters Mooney-Rivlin material was assumed for the simulations : $a_{10}$ = 2500 Pa and $a_{20}$ = 625 Pa.

## 3. Validation procedure

Few authors have proposed extended validation procedures for soft tissue modeling. In maxillofacial surgery, most of them compare their simulations with facial and profile pictures of the patient. While a qualitative comparison is always required, this method is quite inaccurate and does not afford a real tri-dimensional evaluation. The main other approach rely on the acquisition of the post-surgical patient morphology with an optical laser scanner. This enable a 3D quantitative comparison (Koch et al. 1999). However, it is very sensitive to the accuracy of the skin surface and the registration procedure to express it in the pre-operative patient referential. Moreover, there is always an important error between the simulated intervention and the bone repositioning actually realized during the surgery. The most advanced quantitative evaluation was recently proposed by Vandewalle et al. (2003), who measure the distances between their simulations and a post-operative CT scan.

The evaluation protocol we propose also requires the acquisition of a pre and a post-operative. While a post-operative exam is invasive in terms of radiations, it is clearly the best available data to assess the quality of numerical simulations. With the improvement of modern scanners, its use can therefore be acceptable in a research context.

Our evaluation procedure consists in four steps:
1. measuring the bone repositioning actually realized during the surgery, by direct comparison of the pre- and post-operative data;
2. simulating the bone osteotomies and applying the measured displacements to the bone segments;
3. simulating the resulting soft tissue deformation using the biomechanical model;
4. evaluating the differences between the simulation and the post-operative data, both qualitatively and quantitatively.

The first two steps are realized using mathematical tools initially developed for a 3D cephalometry project (Chabanas et al. 2002). Anatomical landmarks in areas that are not modified during the surgery are defined in both the pre- and post-operative CT slices, to register the two datasets in a same referential. Then, landmarks located on each bone segment (e.g. the mandible and maxillae) enable to measure the displacements actually applied during the surgery (figure 1). Although the anatomical landmarks are manually positioned on the CT slices, it has been shown that their repeatability is in mean .25 mm, which yields to very acceptable results in the measurements of the bone displacements.

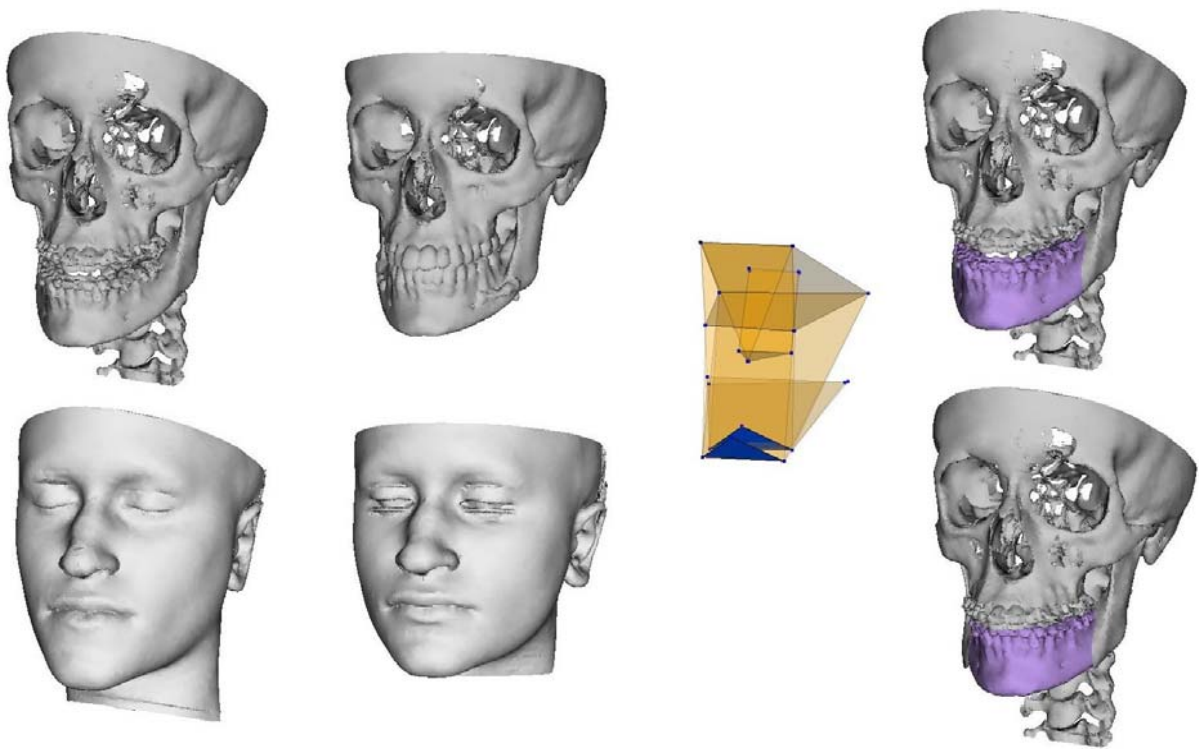

Figure 1 : clinical case of mandibular prognatism. Left, the patient skull and skin surface, before and after the surgery. By comparison of the two skeleton surfaces using our 3D cephalometry, the mandibular correction actually realized during the intervention was accurately measured and reproduced (right).

The measured displacements define the boundary conditions for the Finite Element model. Inner nodes in contact with the non-modified skeleton surface are fixed, while the measured displacements are applied to the nodes on the osteotomized bone segments. Nodes around the osteotomy line are not constrained, to account for the bone-tissue separation due to the surgical access. Rest of the nodes, in the outer part of the mesh or in the mouth and cheeks area are let free to move.

Once the outcome of the surgery has been simulated with the biomechanical model, it can be compared with the post-operative skin surface of the patient, reconstructed from the CT scan. The quantitative comparison between the two datasets is achieved using the MESH software (Aspert et al. 2002), which has been improved to calculate signed Euclidian distances.

## 4. Results

Results are presented on a clinical case of retromandibular correction. A pre-operative CT scan was first acquired, which enabled us to generate a 3D mesh conformed to the patient morphology. After the patient was operated in a conventional way, a post-operative CT scan was acquired. By comparing both datasets, the actual displacement applied to the mandible during the intervention was measured (figure 1). It consisted in a backward translation of 0.9mm (in the mandible axis), and a slight rotation in the axial plane. The measured procedure was then reproduced on the skeleton model, and boundary conditions for the Finite Element model were set.

By comparison of the vertebras positions in both CT scans, it can be seen that the inclination of the head was not similar during both exams, with a difference of more than 10 degrees. Unfortunately, this imply the simulations would not be comparable with the post-operative aspect in the neck area, since the head position directly influence the cervico-mentale angle. Therefore, another boundary conditions was added to the extreme lower nodes of our mesh, in the neck area, to reproduce this modification of the head position. Although quite qualitative, this should enable us to better compare the simulations with the actual patient morphology.

Simulations were computed using the Ansys$^{TM}$ Finite Element software (Ansys Inc.). For the linear elastic model, the computing time is less than 3 seconds with the small deformation hypothesis, and almost 3 minutes in large deformations. The hyperelastic calculus required up to 8 minutes. All simulations were static, and ran on a 2.4 GHz PC. The model counts around 5000 elements and 7650 nodes. For the two non-linear models, numerical convergence can be difficult to obtain if the boundary conditions are not well defined. This was particularly the case in our first trials, before the mesh was extended in the posterior direction, which enabled us to better constrain the posterior nodes. Generally speaking, it must be recall that convergence is very sensitive to the boundary conditions, the quality of the elements shape and the time-steps used during the numerical resolution.

Figure 2 shows the Von-Mises repartition of the strain, calculated for linear model in small deformation. Figure 3 presents the results obtained with the linear elastic model in large deformation, along with the post-operative skin surface of the patient. Finally, figure 4 shows the distances measured between each model and the patient data with the MESH software.

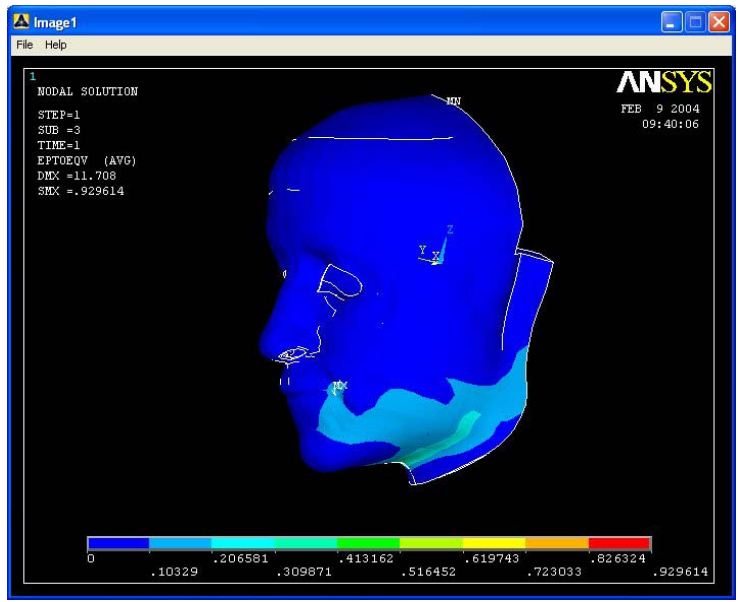

Figure 2 : repartition of the strain for the linear model in small deformation.

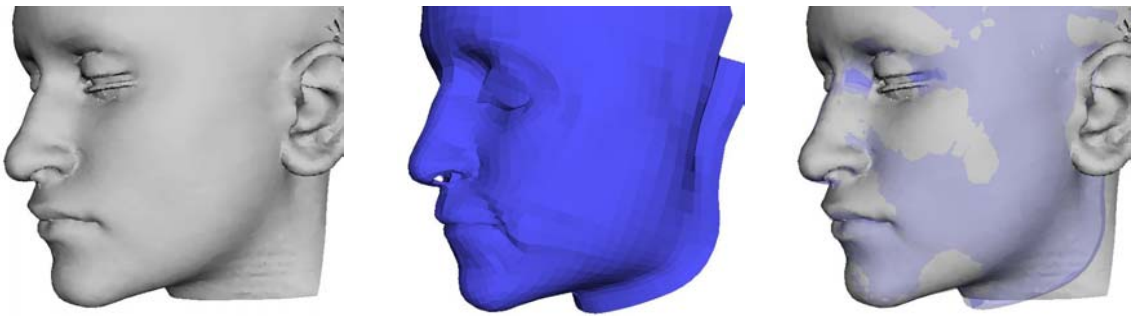

Figure 3: comparison of the post-operative data (left) with the linear elastic model in large deformation (center). Both models are superposed in the right view.

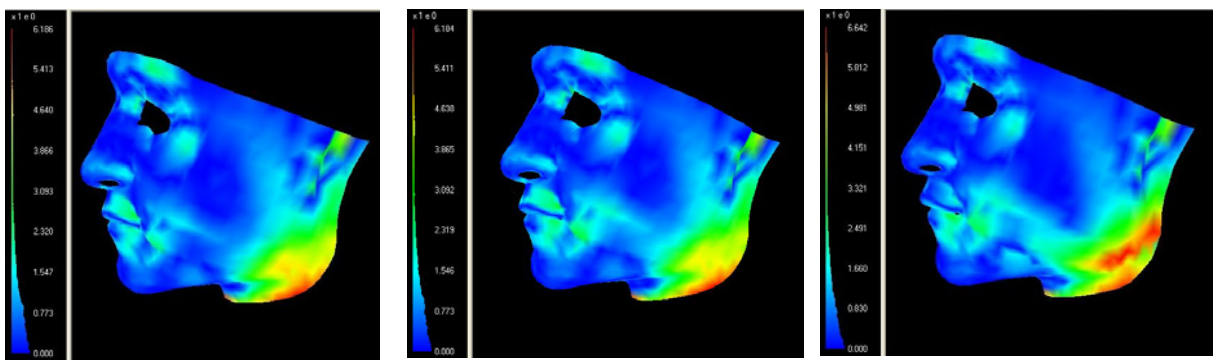

Figure 4: measure of the error between the simulations and the post-operative data, for the linear model in small (left) and large deformation (center), and the hyperelastic model (right).

# 5. Discussion

Before analyzing the results, a methodological point should be discussed. The use of numerical data (CT scan) enable us to obtain a quantitative evaluation of the simulation errors. Such results are quite rare (only Vandewalle et al. (2003) got similar ones with actual post-operative data) and seems extremely important and necessary to really assess the influences of the different modeling options. Nevertheless, the numerical values should be carefully analyzed. They represent the minimal Euclidian distance between points of the model and the patient skin surface, which does not mean the distances to their real corresponding points (e.g. the distance between a point P and the surface S can be smaller than the distance between P and its corresponding points P' in S). These numerical values are thus always a minimization of the true errors. Therefore, the quantitative analysis must always be completed by a qualitative evaluation of the results, carried out by a clinician. This remains the best way to know how well the model is perceived by the surgeon, an gives an emphasis to the most relevant morphological areas in the face: cheeks bones, lips area, chin and mandible angles.

First, it should be noted that the simulations obtained with all models are quite similar. This is an interesting result since the stress repartition (figure 2) shows that the relative stress are above 20% in a large area around the mandible, with peak of 30% to 100% in the osteotomy area (the maximum being in the bone surface region). A first conclusion is then that a linear model, even is small deformation, appears quite acceptable even for relative deformation up to 20%. The large deformation hypothesis does really not decrease the errors. Surprisingly, the first results obtained with the hyperelastic model shows more important errors. This could be explained by the fact this modeling is much more sensitive to several critera like the mesh (which must be refined in the high-stress areas), the boundary conditions and the rheological parameters. Such complicated model require more testing before being used, and may not the most adapted for problems with relatively small deformations.

Clinically, the simulations are of good quality and quite coherent with the actual outcome of the surgery. The accuracy is the best in the chin area, which is logical since that region is one of the most constrained. A slight swelling is observed in the cheeks area of the model, which is a known clinical behavior in retro-mandibular procedures and was correctly reproduced with the model.

Although they are not the largest numerically, less than 2 mm, errors around the lips are important. It can be observed that the shape of the inferior lips is unchanged from the pre-operative state, just moved, and thus incorrect. This is explained by the fact *contacts* between both lips and between the lips and the teeth are not taken into account so far. Indeed, penetration occurred with the teeth, which would certainly have modified the shape of the lip if the contacts were handled. This essential modeling aspect, not discussed in the literature of facial simulation, will really have to be integrated.

The most important errors, up to 6mm, occur around the angles of the mandible. Numerical values should be analyzed carefully since the difference of head inclination certainly influence the results. Nevertheless, errors are expected important in that areas where the stress is maximum. They correspond to the osteotomy region, and thus the frontier between constrained and free nodes. The swelling observed in all models is more important that in the actual data. Before complicating the model, for example with growth modeling (Vandewalle et al. 2003), we prefer to continue the tests and evaluations, since the behavior in these areas appears quite sensitive to the boundary conditions, especially for the two non-linear models.

# 6. Conclusion

A qualitative and quantitative evaluation procedure was proposed in this paper and used to compare different modeling of the face soft tissue. First results are quite encouraging. It has particularly been shown that for the simulation in maxillofacial surgery, a linear elastic model can be sufficient for simple procedures like retro-mandibular correction.

Future works are to extend the evaluation of different modeling options and to assess the influence of elements (refinement, linear or quadratics elements…), rheological properties and numerical methods. Lips and lips-teeth contact must also be taken into account. Two more complex clinical cases are planned for the evaluation (with a post-operative CT scan): a bimaxillary correction with genioplasty, and a distraction of the orbito-zygomatic structure. The non-linear models are expected to be necessary to simulate these difficult, large deformations procedures.